\documentclass[showpacs,floatfix]{revtex4}
\usepackage{graphicx}
\usepackage[dvips]{epsfig}
\usepackage{dcolumn}
\usepackage{subfigure}%
\usepackage{bm}
\usepackage{amssymb}
\usepackage{amsmath}

\begin{document}
\title{Extinction in population dynamics}

\author{C. Escudero$^{\dag}$, J. Buceta$^{\ddag}$, F. J. de la
Rubia$^{\dag}$, and Katja Lindenberg$^{\ddag}$ }

\affiliation{
\dag Departamento de F\'{\i}sica Fundamental,
Universidad Nacional de Educaci\'on a Distancia, C/ Senda del Rey 9,
28040 Madrid, Spain\\
\ddag Department of Chemistry and Biochemistry, and Institute
for Nonlinear Science, University of California San Diego,
9500 Gilman Dr., La Jolla, CA 92093-0340, USA}

\begin{abstract}
We study a generic reaction-diffusion model for single-species population
dynamics that includes reproduction, death, and competition.  The
population is assumed to be confined in a refuge beyond which conditions
are so harsh that they lead to certain extinction.  Standard continuum
mean field models in one dimension yield a critical refuge length $L_c$
such that a population in a refuge larger than this is 
assured survival.  Herein we extend the model to
take into account the discreteness and finiteness of the
population, which leads us to a stochastic description.  We present a
particular critical criterion for likely extinction, namely, that the standard
deviation of the population be equal to the mean.  According to this
criterion, we find that while survival can no longer be guaranteed for
any refuge size, for sufficiently weak competition
one can make the refuge large enough (certainly larger than $L_c$)
to cause extinction to be unlikely.  However, beyond a certain value of
the competition rate parameter it is no longer possible to escape a
likelihood of extinction even in an infinite refuge.  These unavoidable
fluctuations therefore have a severe impact on refuge design issues.
\end{abstract}

\maketitle

\section{Introduction}
\label{sec:introduction}
Population dynamics is a venerable and important subject 
that has been studied thoroughly for decades~\cite{murray}.
Understanding ecological systems is interesting in itself, as are
applications of this understanding to a large variety of important
practical problems such as, for example, the spread of a virus or
other disease over a host population~\cite{abramson}, and strategies for
the elimination of pests or for the protection of endangered
species~\cite{courchamp}.
These examples point to the importance of the study of the extinction of
populations and of the conditions that lead to extinction as well as
those that protect against it. 

The oldest population models pose relatively simple rate equations
for a total population, e.g., those of the logistic variety, and
typically take basic events such as births and deaths into account
through appropriate
rate processes.  A generic example is the logistic
form~\cite{horsthemke,doering}
\begin{equation}
\frac{dX}{dt}=\mu X -X^2,
\label{logistic}
\end{equation}
where $X$ is a measure of the size of a population and is therefore
nonnegative. The parameter $\mu$ is the difference between birth and
death rates of the population, and $X=0,\mu$ are its steady states. All
that is required for survival is that $\mu>0$.
Fairly early on, the possible role of fluctuations in
these models was recognized and included by way of additional
fluctuating terms. The sources of the fluctuations
included in this way in the earliest models were associated with
variations in the external environment in which the population evolves,
and appeared as a fluctuation in the parameters of the equations.  For
example, a fluctuating birth and/or death rate in the logistic form
might lead to the description~\cite{horsthemke}
\begin{equation}
\frac{dX}{dt}=\mu X-X^2+\xi(t)X,
\end{equation}
where $\xi(t)$ is a zero-centered Gaussian white noise.
The role of ``internal fluctuations'' arising from the fact that the
populations are finite and discrete was also recognized, and the
stochastic contributions to the population equations were
obtained from an appropriate expansion of a master
equation~\cite{doering}.
Writing the logistic equation in appropriately rescaled form, 
one would have with $Y= X/\mu$ and
$\tau=\mu t$,
\begin{equation}
\frac{dY}{d\tau} =  Y(1-Y) + \sigma \sqrt{Y(1-Y)}\xi(t).
\end{equation}
The parameter $\sigma$ is also determined by the rates that determine
$\mu$.  Other sources of stochasticity have
also been recognized, namely, those arising from natural catastrophes
and from changes in gene frequencies~\cite{shaffer}.

The per capita growth rate of the population according to the
deterministic logistic model \eqref{logistic} is
\begin{equation}
\frac{1}{X} \frac{dX}{dt} = \mu-X
\end{equation}
which clearly decreases to zero as $X$ approaches the steady state value
$\mu$ but that has a maximum as $X\to 0$.  A point that was observed early on
in the biological ecology literature is that this latter behavior is not
descriptive of real populations in many cases, and that in fact the per
capita growth rate also tends to decrease when the population becomes
very small~\cite{courchamp,allee,odum}. 
Various possible mechanisms have been discussed for this in
the literature, but many of them are based on the recognition that not
only the total population but also its {\em density} may play a crucial
role in population dynamics, particularly in questions related to
extinction. 

More recent population models have recognized the
importance of describing not only the total populations but also its
spatial distribution, thereby getting a handle on density-related
issues. Perhaps the easiest way of dealing with a spatial distribution 
is by way of reaction-diffusion equations, and indeed such studies
in the biological/ecological context have a long 
history~\cite{skellam,britton,cantrell,diekmann,fife,ludwig,okubo}.
Reaction-diffusion equations describe deterministic continuous
densities.  Therefore, whereas they
do deal with some of the density-dependence issues of great importance
in the assessment of extinction probabilities, they do not in general
include the other feature discussed above that is also of great
consequence in these problems, namely, the effect of fluctuations.  These
fluctuations can again arise from a variety of internal and external
sources.  Herein we address the problem of {\em internal} fluctuations.

Populations are composed of discrete numbers of individuals.  That
populations are not continuous is particularly relevant at low densities
and at low total populations.  The importance of these effects have been
recognized and analyzed in various ways in recent years, most
prominently in works that deal with a master equation description of
the population instead of a reaction-diffusion
model~\cite{sznajd1,sznajd2}.  These master
equations are typically solved numerically and provide important
information on various measures for extinction.

We take a different approach to the problem, our interest lying in an
analytic formulation of the problem of internal fluctuations and their
consequences.  On the one hand, an analytic approach necessarily
requires a more narrow focus on a particular question, and therefore
requires some restriction as to the range of issues that we can address. 
On the other hand, analytic results may provide insights that are
difficult to obtain in purely numerical work.

Our specific question arises from the following situation.
Suppose that a population lives in a one-dimensional bounded refuge with
favorable conditions for life, while outside of the refuge the conditions are
extremely harsh.  A question that has been considered prominently in the
mathematical literature is that of the critical size $L_c$ of such a refuge
that will guarantee the survival of the
population~\cite{britton,ludwig,okubo}, that is, the critical size
beyond which the extinction probability vanishes, $P_{ext}=0$. This
problem, and many
generalizations of it, have been solved, but to our knowledge existing
analytic solutions are based on the assumption of a continuous population.
We show that accounting for the discreteness and finiteness of the population
profoundly modifies the known results and may reopen an
array of questions for reconsideration.

Our starting point is a master equation to describe the following
scenario. In a refuge of volume $V$ 
we have a population composed of a single stream of individuals called $A$ 
that diffuse with constant diffusion coefficient $D$ on a
$d$-dimensional lattice (later $d$ is set to unity) with lattice
spacing $\Delta x$. We further
suppose that the individuals reproduce by fission ($A\rightarrow A+A$, as
may be the case with bacteria) with rate constant $\sigma_1$,
they die ($A \rightarrow \emptyset$) with rate coefficient $\sigma_2$,
and they compete according to the reaction $A+A \rightarrow A$
with rate coefficient $\lambda$. We suppose that $\sigma_1>\sigma_2$,
because otherwise the population eventually becomes extinct with certainty.
We also assume that the nutrients are homogeneously distributed so
that the coefficients do not depend on the lattice site. 
The microscopic description of the system within the refuge is given by the
master equation
\begin{eqnarray}
\frac{dP(\{n_i\};t)}{dt}&=&\sum_i \biggl\{\frac{D}{(\Delta x)^{2}}
\sum_{\{m\}}\left[(n_m +1)P(\ldots,n_i-1,n_m +1,\ldots;t)
\right.\nonumber\\
&&\left.~~~~~-n_i P(\ldots,n_i,n_m,\ldots;t)\right] \nonumber\\ 
&&~+\lambda\left[(n_i+1)n_iP(\ldots,n_i+1,\ldots;t)-n_i(n_i-1)
P(\ldots,n_i,\ldots;t)\right]
\nonumber\\
&&~+\sigma_1 \left[(n_i-1)P(\ldots,n_i-1,\ldots;t)-n_i
P(\ldots,n_i,\ldots;t)\right]
\nonumber\\
&&~+ \sigma_2 \left[(n_i+1)P(\ldots,n_i+1,...;t)-n_i
P(\ldots,n_i,\ldots;t)\right] \biggr\}.
\label{master}
\end{eqnarray}
Here the occupation number $n_j$ is the number of bacteria at site $j$ and
$\{n_j\}\equiv(\ldots, n_{j-1},n_j,n_{j+1},\ldots)$ is the set of all 
occupation numbers.  The index $i$ in the sum
ranges over all lattice sites and ${m}$ denotes the set of nearest
neighbors of $i$. We choose a homogeneous initial 
condition given by an uncorrelated Poisson distribution
with average population ${\mathfrak n}$ at each site:
\begin{equation}
\label{initial}
P(\{n_i\};0)=e^{-{\mathfrak n}}\prod_i \frac{{\mathfrak n}^{n_i}}{n_i!}.
\end{equation}
Outside of the refuge living conditions are assumed to be extremely harsh,
represented by the addition of the process $A \rightarrow \emptyset$
with a large rate coefficient $\gamma$.  In the limit
$\gamma\rightarrow\infty$ (certain death outside of the refuge), we can
implement this contribution by requiring that $P(\{n_i\};t)$ vanish
whenever $n_k \neq 0$, where $k$ denotes any site just beyond the refuge
boundary.  Thus, ``harshness'' is implemented as a set of boundary
conditions.

In Sec.~\ref{sec:model} we 
discuss the derivation of a stochastic Langevin equation whose
moments can be related to those of the fluctuating population described 
by the master equation. In
Sec.~\ref{sec:extinction} we present criteria for survival or
extinction.  In this section we discuss whether a refuge can be
made sufficiently
large to make extinction unlikely. A summary of our results and some
directions for future research are presented in
Sec.~\ref{sec:conclusions}.

\section{The Model}
\label{sec:model}

It is well known that the continuum (mean field) description of the
local concentration $\rho({\bm r},t)$ is the
Fisher-Kolmogorov-Petrovsky-Piscunov~(FKPP) equation~\cite{FKPP}
\begin{equation}
\label{deterministic}
\frac{\partial \rho({\bm r},t)}{\partial t}=D\nabla^2
\rho({\bm r},t) +(\sigma_1-\sigma_2) \rho({\bm r},t)-
\lambda \rho({\bm r},t)^2,
\end{equation}
which has been the phenomenological starting point in a huge number of
problems in physics, chemistry, and biology.  In one dimension and with
the refuge extending inside the interval
$\left[ -\frac{L}{2},\frac{L}{2} \right]$, the equation is to be solved
subject to the boundary conditions 
\begin{equation}
\rho \left( -\frac{L}{2},t \right)=\rho \left( \frac{L}{2},t \right) =0.
\label{boundary}
\end{equation}
For small $L$ the only solution to this problem
as $t\rightarrow\infty$ is $\rho(x,t) \rightarrow 0$, that is, the
population becomes extinct.  There is a critical refuge size $L_c$
beyond which extinction does not occur and the population will be
nonzero (albeit small). Since $\rho(x,t)$ is small near $L_c$, to
find this critical value it is appropriate to linearize
Eq.~\eqref{deterministic} around zero population density
and solve the simpler equation~\cite{ludwig}
\begin{equation}
\label{classic}
\frac{\partial \rho(x,t)}{\partial t}=D\frac{\partial^2
\rho(x,t)}{\partial x^2} +(\sigma_1-\sigma_2) \rho(x,t), 
\end{equation}
subject to the boundary conditions (\ref{boundary}).
The Fourier decomposition of the solution is given by
\begin{eqnarray}
\rho(x,t)&=&\sum_n a_n\rho_n(x,t), \nonumber\\
\rho_n(x,t)&=&\exp \left[ \left( \sigma -\frac{n^2 \pi^2}{L^2}D
\right)t \right]
\sin \left[ \frac{n \pi}{L} \left(x + \frac{L}{2} \right)
\right],
\end{eqnarray}
where $\sigma=\sigma_1-\sigma_2$, $n=1,2,...$, and the $a_n$ depend on
the initial condition.
Thus, all Fourier modes decrease to zero in time if and only if
$L<\pi \sqrt{\frac{D}{\sigma}}$, leading to the conclusion that for
these values of $L$, $\rho_n(x,t) \rightarrow 0$
when $t \rightarrow \infty$ for any $n$. We can therefore conclude that
the critical length of the refuge is
\begin{equation}
L_c=\pi \sqrt{\frac{D}{\sigma}}.
\end{equation}
That is, the population becomes extinct with certainty if
$L<L_c$ and it certainly does not become extinct if $L>L_c$ (in fact, the
Fourier components grow without bound in this case, but the linearized
equation is then no longer valid). 
This result, and extensions of it, have been known for more than fifty
years~\cite{skellam}, and has served as a background for the
design and analysis of refuges.  Note that the linearization argument
that leads to Eq.~\eqref{classic}
is valid for any nonlinearity that can be neglected near extinction in
Eq.~\eqref{deterministic}, and that therefore the resulting $L_c$ is
obtained not only for this specific model~\cite{ludwig}.

The above mean field analysis does not take into account the effect of
the fluctuations associated with the fact that the population is really
discrete and finite.  To do this short of solving the full master
equation, one must formulate a generalization of the FKPP model that
takes the resulting fluctuations into account.
Such generalizations exist in the literature in other contexts.  In
particular, on the basis of a theory first proposed by Doi~\cite{doi},
further elucidated by Peliti~\cite{peliti}, used for the study of critical
phenomena associated with bulk transitions in reaction
dynamics~\cite{cardy}, explicitly applied to a
reaction-diffusion front problem
by Pechekin and Levine~\cite{levine}, and subsequently used in
a number of other contexts related to Fisher waves (see e.g.~\cite{moro}),
one arrives at a stochastic differential equation for a field whose 
moments can be related to those of the
population in the original master equation.  The connection comes about as
follows.  The master equation Eq.~\eqref{master} can be projected onto a
problem in quantum field
theory by the construction of an adequate Fock space. Since this
calculation has been performed many times in the literature we 
only indicate some of the main steps.
A detailed review of the procedure can be found in~\cite{cardy}. 
Consider the following operator algebra:
\begin{equation}
\label{commutation}
[a_i,a_j^+]=\delta_{ij},\qquad [a_i,a_j]=0,\qquad [a_i^+,a_j^+]=0,
\end{equation}
where $a_i$, $a_i^+$ are destruction and creation
operators and the square brackets denote the commutator~\cite{peliti}.
A state vector can be defined as
\begin{equation}
\label{state}
\left| \psi(t) \right>=
\sum_{\{n_i\}} P(n_1,n_2,...;t)a_1^{+n_1}a_2^{+n_2}...\left| 0 \right>,
\end{equation}
where $P(n_1,n_2,...;t)=P(\{n_i\},t)$ is the solution of the master
equation, and
the sum is performed over all possible configurations of the
$\{n_i\}$.  The state vector obeys the imaginary
time Schr\"{o}dinger equation
\begin{equation}
\label{schrodinger}
\frac{d \left|\psi(t) \right>}{dt}=-\hat{H} \left| \psi(t) \right>
\end{equation}
whose formal solution is
\begin{equation}
\left| \psi(t) \right>=e^{-\hat{H}t} \left| \psi(0) \right>.
\end{equation}
In our case the (non-hermitian) Hamiltonian is 
\begin{equation}
\label{hamiltonian}
\hat{H}=\sum_i\left[-\frac{D}{(\Delta x)^2} \sum_{\{m\}}a_i^+(a_m-a_i)
-\lambda(1-a_i^+)a_i^+a_i^2
+\sigma_1[1-a_i^+]a_i^+a_i-\sigma_2(1-a_i^+)a_i\right].
\end{equation}
Note that we recover Eq.~(\ref{master}) by substituting Eqs.~(\ref{state})
and (\ref{hamiltonian}) into Eq.~(\ref{schrodinger}).

One introduces the Glauber state as the projection state
\begin{equation}
\left< S \right|=\left< 0 \right| \prod_i^N e^{a_i}.
\end{equation}
The expected values of 
observables $A(\{n_i\})$ can then be written as
\begin{equation}
\left< A(t) \right> = \sum_{\{n_i\}} A(\{n_i\})P(\{n_i\};t)
= \left<S \right| \hat{A}e^{-\hat{H}t} \left|\psi(0)\right>,
\end{equation}
where $\hat{A}$ is the operator obtained by replacing every $n_i$ in the
function $A(\{n_i\})$ by the number operator $\hat{n}_i=a_i^+a_i$.

The steps that we do not repeat~\cite{cardy,levine} show that this
second-quantized theory can be expressed in terms of a path integral.
Furthermore, from this path integral one can derive a ``classical'' action
from which one can in turn derive the mean field continuum equation
Eq.~\eqref{deterministic}. 
If one concentrates on the path integral itself instead of moving on to the
classical action, one can show an equivalence between the master equation
and the following Langevin equation~\cite{cardy,levine}:
\begin{eqnarray}
\frac {\partial \psi({\bm r},t)}{\partial t}&=&D\nabla^2\psi({\bm r},t)
+(\sigma_1-\sigma_2) \psi({\bm r},t)-\lambda \psi^2({\bm r},t) 
\nonumber\\ \nonumber\\
&& + \sqrt{2\left[ \sigma_1\psi({\bm r},t) -\lambda \psi^2({\bm
r},t)\right]} \xi({\bm r},t),
\label{stochastic}
\end{eqnarray}
where $\xi({\bm r},t)$ is Gaussian white noise with mean and correlation
given by~\cite{cardy2}:
\begin{equation}
\left< \xi({\bm r},t) \right>=0,
\end{equation}
\begin{equation}
\left< \xi({\bm r},t) \xi({\bm r}',t') \right>= \delta(t-t')
\delta({\bm r} -{\bm r}').
\end{equation}
The multiplicative noise in
Eq.~\eqref{stochastic} must be interpreted according to It$\hat{\rm o}$.

It is important to note that 
the noise in Eq.~\eqref{stochastic} can be imaginary, so the value of
$\psi$ is in general complex.  This makes it clear that the field
is {\em not} to be interpreted as the population density.  The field
$\psi$ is the complex eigenvalue of the destruction operator, so
one can relate its moments to the 
moments of the population density using the commutation
relations Eq.~(\ref{commutation}) and the property of the Glauber state 
$\left< S \right| a^+= \left< S \right|$.  For the first two moments we have:
\begin{eqnarray}
\left< \rho({\bm r},t) \right>=\left< n_i \right>=\left< a_i^+a_i \right>=
\left< S \right| a_i^+a_i \left| \psi(t) \right>=
\left< S \right| a_i \left| \psi(t) \right>=
\left< a_i \right>=\left< \psi({\bm r},t) \right>, \nonumber \\
\left< \rho^2({\bm r},t) \right>=\left< n_in_i \right>=
\left< a_i^+a_ia_i^+a_i \right>=
\left< a_i^+a_i^+a_ia_i+a_i^+a_i \right>= 
\left< S \right| a_i^{+2}a_i^2+a_i^+a_i \left| \psi(t) \right> \nonumber \\
=\left< S \right| a_i^2 \left| \psi(t) \right> + \left< S \right|
a_i \left| \psi(t) \right>= \left< a_i^2 \right>+ 
\left< a_i \right>= \left< \psi^2({\bm r},t) \right> +
\left< \psi({\bm r},t) \right>.
\label{mean}
\end{eqnarray}
Higher moments can also be calculated following this procedure.  Here
the lattice site $i$ is associated with the volume $(\Delta x)^d$ around
point ${\bm r}$ and we have set $\Delta x \equiv 1$ for economy of
notation.

As a side note we point out that the so-called stochastic FKPP equation
of the form
\begin{equation}
\label{classic2}
\frac{\partial \rho(x,t)}{\partial t}=D\frac{\partial^2
\rho(x,t)}{\partial x^2} +\sigma_1 \rho(x,t) -\lambda \rho^2(x,t)
+\sqrt{2\left[ \sigma_1\rho(x,t) -\lambda \rho^2(x,t)\right]} \xi(x,t)
\end{equation}
can not represent the population density either since it does not allow
fluctuations around the equilibrium population density
$\rho=\sigma_1/\lambda$ because the population is real, while
these fluctuations do not vanish
in the original discrete system.  However, a connection between
Eq.~\eqref{classic2} and the original master equation Eq.~\eqref{master}
without the death process has
recently been elaborated~\cite{doering}.  This connection is different
from the one
presented above involving the stochastic equation (\ref{stochastic}).

\section{Extinction Probability}
\label{sec:extinction}

The mean field model for the refuge population predicts certain extinction
if the refuge size $L<L_c$ and certain survival if $L>L_c$.  In this
section we show that the fluctuations destroy this certainty of survival,
that is, that survival is in fact never guaranteed.  

In the biological literature one encounters the concept of a {\em
minimum viable population}~\cite{shaffer,sznajd1}.  There are some
variations in its
definition, but one that seems to be widely accepted is that
a minimum viable population is the smallest isolated population having a
99\% chance of remaining extant for 1000 years despite the presence of
foreseeable fluctuations.   Clearly one can vary the two numbers that
appear in this criterion, but, in any case, it involves a {\em probability
of survival} in a {\em finite period of time}.  We wish to make clear at
the outset that our analytic theory is not yet developed to the point of
being able to handle finite times, and we are therefore not able to
implement a criterion of this sort.  Instead, the theory (like the
mean field prediction stated above) deals only with steady state
probabilities.  

Near extinction one can show directly from the master equation
\eqref{master} that the competition term (the term proportional to
$\lambda$) can be neglected.  As a result,
in this limit the contributions 
$\lambda\psi^2$ in the stochastic equation \eqref{stochastic} can
also be neglected, and to find the critical refuge size
one only needs to deal with the simpler Langevin equation
\begin{equation}
\frac {\partial \psi(x,t)}{\partial t}=D\nabla^2\psi(x,t)
+\sigma \psi(x,t)
+ \sqrt{2\sigma_1\psi(x,t)} \xi(x,t).
\label{stochastic2}
\end{equation}
The mean value of $\psi$ obeys the equation
\begin{equation}
\frac{\partial \left< \psi(x,t) \right>}{\partial t}=
D\frac{\partial^2 \left<\psi(x,t)\right>}{\partial x^2}
+ \sigma \left< \psi(x,t) \right>,
\end{equation}
which is the same equation, Eq.~\eqref{classic}, that
we solved for the deterministic case.
Since $\langle \psi(x,t) \rangle=\langle \rho(x,t) \rangle$, we have the same
boundary conditions as in Eq.(\ref{classic}), which implies that the
mean value of the population density is zero, that is, that extinction
is certain, for $L<L_c$.  Since $\left< \psi(x,t) \right>=\left< \rho(x,t)
\right> >0$ for $L>L_c$, this in turn implies that 
extinction is certain if and only if $L<L_c$.  This, however, does not
tell us whether or not the extinction probability is zero for $L > L_c$
(as predicted in mean field theory).  Indeed, we now show that
the extinction probability is greater than zero for any finite length
of the refuge.

It is fairly obvious that $P_{ext}$ does not vanish if one has a
finite population and thus a nonzero probability (albeit perhaps small)
of a total extinction event. We will place this
statement on a more formal footing.  However, an arbitrarily small
extinction probability is not necessarily very important since in most
design problems one deals with at least some uncertainties.  It is
therefore useful to establish a criterion as to what constitutes a
non-negligible extinction probability and what the minimal
size $L^\ast$ of the refuge
must be to insure a probability smaller than this.  Clearly, $L^\ast
\geq L_c$.  We will establish
such a criterion and show that (depending on parameter values) the
size $L^\ast$ may be much larger than $L_c$ and even infinite.

First, to establish the obvious fact that fluctuations lead to a nonzero
extinction probability even for $L>L_c$, let us suppose that we begin
with an infinite refuge and exclude death events $A \rightarrow
\emptyset$ altogether.
The population inside an interval of length $L$ of this infinite refuge
is then greater than or equal to the population inside our actual finite
refuge of length $L$.
The population density inside the finite interval of the infinite refuge in
the steady state is given by the Poisson distribution
\begin{equation}
P(\{n_i\})=e^{-2\sigma_1/\lambda}\prod_i \frac{(2\sigma_1/
\lambda)^{n_i}}{n_i!},
\end{equation}
so that the probability of an unbounded realization of the process is
zero. Since the integral of a bounded function over
a bounded interval is bounded, we can conclude that the
total population inside the interval is finite. Therefore so is the
total population inside the finite refuge.  Furthermore,
with $\delta$-correlated Gaussian fluctuations~\cite{mfpt1,mfpt2},
even the most rare fluctuations occur in a {\em finite} mean time
and therefore
a rare fluctuation caused by the reaction $A \rightarrow \emptyset$ could
kill the entire population in a finite time. We thus conclude that
the probability of extinction is greater than zero whenever $L<\infty$.
In the case
of an unbounded refuge the mean value of the population density is
$\sigma/ \lambda$ and is homogeneously distributed, so a local fluctuation
cannot kill the population. We can summarize these conclusions as follows:
\begin{eqnarray}
P_{ext}=1 \hspace{1 cm} \mathrm{if} \hspace{1 cm} L<L_c \nonumber\\
0<P_{ext}<1 \hspace{1 cm} \mathrm{if} \hspace{1 cm} L_c<L<\infty \nonumber\\
P_{ext} \rightarrow 0 \hspace{1 cm} \mathrm{when} \hspace{1 cm}
L \rightarrow \infty .
\end{eqnarray}
We can furthermore state that since the number of individuals
in the population increases as the length of the refuge increases, the 
probability of extinction decreases monotonically and continuously
as $L$ increases continuously.
Thus, consideration of the internal fluctuations drastically changes
the picture; now there is no finite size of the refuge that
can be considered absolutely safe for the population. However, we have
not been able to derive the explicit functional dependence of $P_{ext}$
on the size of the refuge.

As an alternative to such an explicit full solution,
we propose a criterion for deciding
when a population is under substantial risk of extinction, and calculate the 
critical size $L^\ast$ of the refuge associated with this criterion.
Suppose that a population density $\rho(x)$ in a refuge of
length $L$ described
according to mean field Eq.~\eqref{deterministic} has a maximum steady state
density $M>0$.  Clearly, $L\geq L_c$.  The maximum occurs at $x=0$, in the
middle of the refuge, that is, $\rho(x=0)=M$~\cite{ludwig,gidas}. 
Now consider the description
of this population that includes
the fluctuations, and compare the average population density
$\langle\psi(x)\rangle =
\langle\rho(x)\rangle$ with the deterministic result $\rho(x)$. 
This average distribution also has a maximum at $x=0$ (see below),
$\langle \psi(x=0)\rangle=M'$.
We define the critical size $L^\ast$ as the refuge length associated
with two conditions:
\begin{enumerate}
\item
The deterministic and stochastically obtained
maxima are equal, $M=M'$; 
\item
The standard deviation of the stochastic population density
is equal to its mean at
each $x$.
\end{enumerate}
Clearly, our idea is that a population density whose standard deviation
everywhere equals its
mean is in danger of extinction, and we call this a ``critical
population.'' It is not clear {\em a priori} that these conditions can be
simultaneously satisfied, but in fact it turns out that they can.
The second condition above says that for the critical population
\begin{equation}
\left<\rho(x,t)\right>^2 = \left<\rho^2(x,t)\right>-
\left<\rho(x,t)\right>^2,
\end{equation}
or, equivalently, in terms of the field $\psi$, one for which
\begin{equation}
\left<\psi(x,t)\right>^2 = \left<\psi^2(x,t)\right>+\left<\psi(x,t)\right>
-\left<\psi(x,t)\right>^2. 
\label{critical}
\end{equation}
From Eq.~(\ref{stochastic}) we obtain the following equation
linking the first and second moments of $\psi$: 
\begin{equation}
\frac{\partial \left<\psi(x,t)\right>}{\partial t}=D\nabla^2
\langle\psi(x,t)\rangle+\sigma \langle\psi(x,t)\rangle-\lambda
\langle\psi(x,t)^2\rangle.
\end{equation}
Inserting Eq.~\eqref{critical} for the critical field then results in a
closed equation for the first moment,
\begin{equation}
\label{criticalfield}
\frac{\partial \langle\psi(x,t)\rangle}{\partial t}=D\nabla^2
\langle\psi(x,t)\rangle+(\sigma +\lambda )\langle\psi(x,t)\rangle
-2\lambda \langle\psi(x,t)\rangle^2.
\end{equation}
Note that the apparent irrelevance of the fluctuations in
Eq.~\eqref{stochastic} other than their mean value in arriving at
Eq.~\eqref{criticalfield} is illusory since the relation \eqref{mean}
and consequently Eq.~\eqref{critical} are intimately connected to the
precise form of the fluctuations.

In the steady state Eq.~\eqref{criticalfield} reduces to the boundary
value problem
\begin{equation}
\label{steady}
D\frac{d^2 \langle\psi(x)\rangle}{dx^2} +
(\sigma+\lambda) \langle\psi(x)\rangle - 2 \lambda \langle \psi(x)\rangle^2
= 0,
\end{equation}
with boundary conditions as in Eq.(\ref{classic}).  
To find $L^\ast$ we multiply Eq.~\eqref{steady} by
$d \langle\psi(x)\rangle/dx$ and rewrite the result as 
\begin{equation}
\frac{d}{dx}\left[\frac{D}{2}\left(\frac{d \langle\psi(x)\rangle}{dx}\right)^2
+ \frac{(\sigma + \lambda)}{2} \langle\psi(x)\rangle^2 -
\frac{2\lambda}{3}\langle\psi(x)\rangle^3\right]=0,
\end{equation}
from which it follows that
\begin{equation}
\frac{D}{2}\left(\frac{d \langle\psi(x)\rangle}{dx}\right)^2
+ \frac{(\sigma + \lambda)}{2} \langle\psi(x)\rangle^2
- \frac{2\lambda}{3} \langle\psi(x)\rangle^3 =  {\rm constant}. 
\end{equation}
Symmetry considerations show that the solution of this 
problem has only one maximum at $x=0$~\cite{gidas}, and since the
first derivative of the solution should vanish at any maximum, 
we can explicitly write the constant as
\begin{equation}
\label{firstintegral}
\frac{D}{2}\left(\frac{d \langle\psi(x)\rangle}{dx}\right)^2 +
\frac{\sigma + \lambda}{2} \langle\psi(x)\rangle^2
- \frac{2\lambda}{3} \langle\psi(x)\rangle^3
=\frac{\sigma + \lambda}{2} M^2 - \frac{2\lambda}{3} M^3,
\end{equation}
where $M$ is the maximum introduced earlier. Since this
maximum occurs at
$x=0$, we can integrate Eq.~(\ref{firstintegral}) 
for $x>0$ to obtain
\begin{equation}
x=\sqrt{\frac{D}{2}} \int_{\langle\psi(x)\rangle}^M
\frac{dz}{\sqrt{\frac{\sigma +
\lambda}{2} M^2 - \frac{2\lambda}{3} M^3
-\frac{\sigma+\lambda}{2} z^2 + \frac{2\lambda}{3} z^3 }}.
\end{equation}
Taking into account that the population vanishes at the edge of the
refuge we can obtain the value of the length $L$ of the refuge as
a function of the maximum of the average population density:
\begin{equation}
L=\sqrt{2D} \int_0^M \frac{dz}{\sqrt{\frac{\sigma+\lambda}{2} M^2 - \frac{2\lambda}{3} M^3-\frac{\sigma + \lambda}{2} z^2 +
 \frac{2\lambda}{3} z^3 }}.
\end{equation}
With the change of variables $w=z/M$ and setting
$\varepsilon\equiv \lambda/\sigma$
we can rewrite this relation as
\begin{equation}
\sqrt{\frac{\sigma}{D}}L=2\sqrt{\frac{1}{1 + \varepsilon}}
\int_0^1 \frac{dw}{\sqrt{1 - w^2 -
\frac{4}{3}\frac{\varepsilon}{(1 + \varepsilon)}(1- w^3) M}}.
\label{l1}
\end{equation}
The steady state deterministic equation obtained by setting the left hand
side of Eq.~\eqref{deterministic} equal to zero is formally
identical to the critical field equation Eq.~\eqref{steady} with modified
parameters.  Since we are choosing $L^\ast$ as the value where the maximum
of the critical average population density is equal to the maximum of the
mean field population density, the mean field model provides us with a second
relation between $M$ and $L$:
\begin{equation}
\sqrt{\frac{\sigma}{D}}L=2\sqrt{\frac{1}{\varepsilon}}
\int_0^1 \frac{dw}{\sqrt{1 - w^2 -
\frac{2}{3}\varepsilon(1- w^3) M}}.
\label{l2}
\end{equation}
The simultaneous solution of Eqs.~\eqref{l1} and \eqref{l2} then leads to
the value of $L^\ast$ (and incidentally also of $M$).  

\begin{figure}
\begin{center}
\psfig{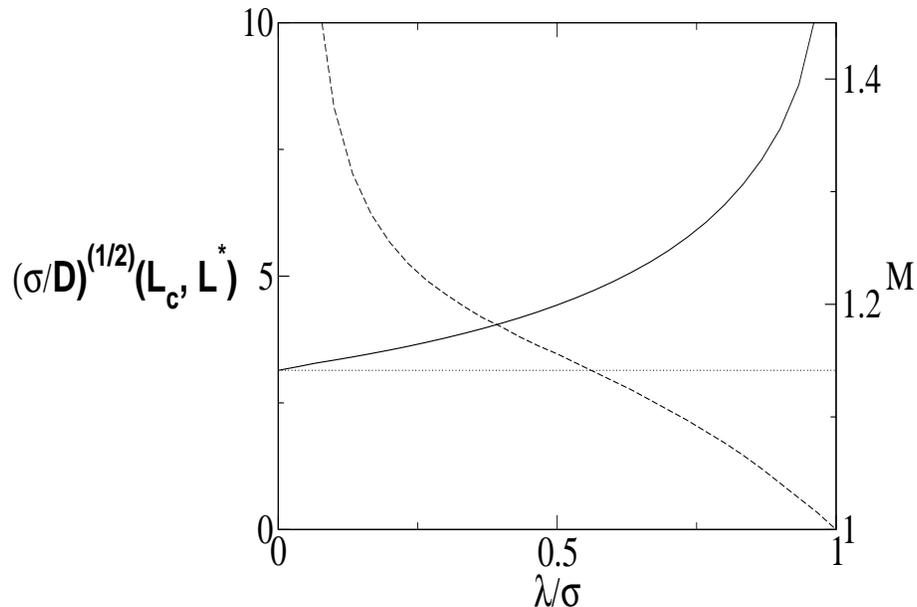} 
\end{center}
\caption{
\label{figura_1} Left scale:
$\sqrt{\sigma/D}L_c$ (dotted line) and $\sqrt{\sigma/D}L^*$ (solid line)
vs $\lambda/\sigma$.
Note that $\sqrt{\sigma/D}L_c=\pi$ for all values of 
$\lambda/\sigma$, while $L^*$ grows monotonically up to $\lambda/\sigma = 1$,
where it diverges logarithmically. Note also
that $L_c=L^*$ when $\lambda = 0$,
and that this is the only common point of the lengths.  Right scale and
dashed curve: maximum
$M$ of the average population density as defined in the text
vs $\lambda/\sigma$.
}
\end{figure}

The two equations reduce to the same equation if $\lambda=0$
($\varepsilon=0$), and in
this case $L^*=L_c$ and $M$ diverges . The reason for the equality
is that when the
nonlinearity is not present, the total steady state population is
unbounded (thermodynamic limit) and
there are no fluctuations.  In general, the two simultaneous
equations can only be solved numerically, and the results are shown in
Fig.\ref{figura_1}.  

A number of important points should be noted about our results.  First,
$L^*$ is indeed greater than $L_c$ except for the single
trivially coincident point
when $\lambda=0$, that is, when there is no competition.  Thus, according
to our criterion whereby the risk of
extinction is substantial when the mean and standard deviation of the
population density are equal, one must increase the size of the
refuge considerably
beyond the deterministic refuge size to avoid this risk. 
Indeed, when $\lambda/\sigma=1$ the size $L^*$ diverges logarithmically
and the population density becomes uniform at density $\sigma/\lambda$.
As $\lambda$ grows beyond $\sigma$ 
it is no longer possible to avoid considerable risk of extinction for any
finite size of the refuge.  The increase in $M$ with decreasing $L^*$
can be understood by moving from right to left in the figure: decreasing
critical refuge size is associated with the requirement of a larger
population (and hence a larger maximum population density) to insure against
extinction.  At $L_c$ the maximum $M$ diverges because no
population, no matter how large, can be protected against extinction
below this length.
We can obtain an analytic expression for $L^*$
for small values of $\lambda/\sigma$:
\begin{equation}
\sqrt{\frac{\sigma}{D}}L^*=\pi \left[ 1+\frac{\lambda}{2\sigma}+o
\left(\left(\frac{\lambda}{\sigma}\right)^2 \right)\right].
\end{equation}
We can also see analytically that $L^*$ must diverge when
$\lambda/\sigma=1$: at this point the two integrals for $L$ can only be
equal if they both vanish (which they do not) or if they both diverge
(which they do when $M=1$).  

\section{Conclusions}
\label{sec:conclusions}
The main point of this paper has been to assess the effects of the
inevitable internal fluctuations on the prediction of the risk of extinction
of a population in a refuge as a function of the size of the refuge.  In
the usual mean field deterministic model in one dimension, the population
is treated as a continuum and one obtains a critical refuge length $L_c$
such that the population becomes extinct with certainty if the length of
the refuge is below this critical value, $L<L_c$, whereas survival is
certain if $L>L_c$. It is well known that, as predicted by the mean field
approach, the smaller the area occupied by a population, the lower are
its chances for survival~\cite{macarthur}.  However, it is also known
that a small size or low density of a population may also increase its
chances of extinction, a feature that mean field theories do not
capture~\cite{courchamp}.  Our approach deals with both of these
features simultaneously through the interplay of discreteness and
nonlinearity in a spatially distributed model.
We have argued that the discreteness of the population
and its finite size make fluctuations unavoidable, and have shown that
whereas $L<L_c$ still guarantees extinction, $L>L_c$ by no means
guarantees survival.  While we have not been able to find an analytic
expression for the survival probability as a function of $L$, we have 
focused on a sensible risk criterion for which we have found
explicit results up to quadrature.  Comparing deterministic and
average stochastic population densities with a given maximum $M$,
we have chosen to define a new critical length $L^*$ as one for which
the mean of the
population density and its standard deviation are equal, reasoning
that this
variability implies considerable susceptibility of the population to
extinction.  We have shown that $L^*\geq L_c$ and, most importantly,
that $L^*$ diverges when the competition rate coefficient grows beyond
that of the net growth rate of the population.  For some parameter
values it is possible to protect (albeit not with certainty)
a population from extinction by placing
it in a sufficiently large refuge, one that is certainly larger
than that predicted by the standard mean field 
model.  For other parameter values (in particular, when competition is
too strong), it is not possible to evade the risk of extinction
(at least according to our criterion) by enlarging the size of the
refuge.  Clearly, these results have serious implications for the
expectations in the design of refuges.

Many questions still remain to be answered. For instance, an 
exact result for the extinction probability would clarify many issues.
So would the ability to obtain time dependent solutions so as to deal
with a more realistic criterion of survival over a long but finite time
interval.
Also, we have only considered the simplest most generic situation,
whereas real systems (particularly ones not designed in the laboratory)
are likely to be seriously affected by many
complicating factors, some connected with the necessary extension to
higher dimensions. Examples include the presence of
convection, different boundary conditions, and spatial inhomogeneities
inside the refuge.  Similar questions can and should be posed when multiple
species are present.  In any case, in light of our results
it would seem prudent to reconsider other critical size population problems 
to assess the effects of discreteness.
As we have shown in the simple model considered here, the consequences
can indeed be profound.

\section*{Acknowledgments}

The authors gratefully acknowledge input from J. L. Cardy.
C. Escudero is grateful to the Department of Chemistry and Biochemistry
of the University of California, San Diego for its hospitality. This work
has been partially supported by the Engineering Research Program of
the Office of Basic Energy Sciences at the U. S. Department of Energy
under Grant No. DE-FG03-86ER13606,
the Ministerio de Educaci\'{o}n y Cultura (Spain)
through grants No. AP2001-2598 and EX2001-02880680,
and by the Ministerio de Ciencia y Tecnolog\'{\i}a (Spain),
Project No. BFM2001-0291.


\begin{thebibliography} {99}

\bibitem{murray}  J. D. Murray, {\it Mathematical Biology},
2nd ed. (Springer, New York, 1993).

\bibitem{abramson} G. Abramson and V. M. Kenkre, Phys. Rev. E
{\bf 66}, 011912 (2002).

\bibitem{courchamp} F. Courchamp, T. Clutton-Brock, and B. Grenfell,
Trends Ecol. Evol. {\bf 14}, 405 (1999).

\bibitem{horsthemke} W. Horsthemke and R. Lefever, {\em Noise-Induced
Transitions: Theory and Applications in Physics, Chemistry, and
Biology} (Springer Verlag, Berlin, 1984).

\bibitem{doering}
C. R. Doering, C. Mueller, and P. Smereka, Physica A {\bf 325}, 243 (2003)
and references therein;
{\em ibid}, in {\em AIP Conference Proceedings} {\bf 665}, 223 (2003).

\bibitem{shaffer} M. L. Shaffer, BioScience {\bf 31}, 131 (1981).

\bibitem{allee} W. C. Allee {\em et al.}, {\em Principles of Animal
Ecology} (Saunders, 1949).

\bibitem{odum} P. E. Odum, {\em Fundamentals of Ecology} (Saunders,
1959).

\bibitem{skellam} J. G. Skellam, Biometrika {\bf 38}, 196-218 (1951).

\bibitem{britton} N. F. Briton, {\it Reaction-Diffusion Equations and Their
Applications to Biology}, Academic Press, New York (1986). 

\bibitem{cantrell} R. S. Cantrell and C. Cosner, J. Math. Biol.,
{\bf 29}, 315-338 (1991). 

\bibitem{diekmann} O. Diekmann and N. M. Temme, {\it Nonlinear Diffusion 
Problems}, Mathematisch Centrum, Amsterdam, (1976). 

\bibitem{fife} P. C. Fife, {\it Mathematical Aspects of Reacting and 
Diffusing Systems}, Lecture Notes in Biomathematics {\bf 28},
Springer-Verlag, New-York, (1979).

\bibitem{ludwig} D. Ludwig, D. G. Aronson and H. F. Weinberger, J. Math. Biol., {\bf 8}, 217-258 (1979).

\bibitem{okubo} A. Okubo, {\it Diffusion and Ecological Problems:
Mathematical Models}, Biomathematics {\bf 10}, Springer-Verlag, Berlin, (1980).

\bibitem{sznajd1} K. Sznajd-Weron, Eur. Phys. J. B {\bf 16}, 183 (2000).

\bibitem{sznajd2} K. Sznajd-Weron and M. Wola\'{n}ski, Eur. Phys. J. B
{\bf 25}, 253 (2002).

\bibitem{FKPP} R. A. Fisher, Ann. Eugenics {\bf 7}, 355 (1936); A.
Kolmogorov, I. Petrovsky, and N. Piscunov, Mosc. Univ. Bull. Math. A {\bf
1}, 1 (1937).


\bibitem{doi} M. Doi, J. Phys. A {\bf 9}, 1479 (1976).

\bibitem{peliti} L. Peliti, J. Phys. (Paris) {\bf 46}, 1469 (1985).

\bibitem{cardy} J. L. Cardy and U. C. T\"auber, J. Stat. Phys. {\bf 90},
1 (1998), and references therein.

\bibitem{levine} L. Pechenik and H. Levine, Phys. Rev. E {\bf 59}, 3893
(1999).

\bibitem{moro} E. Moro, Phys. Rev. Lett. {\bf 87}, 238303 (2001).

\bibitem{cardy2} The effects of some correlated initial
conditions are discussed in J. Cardy and P-A. Rey, J. Phys. A
{\bf 32}, 1585 (1999).


\bibitem{mfpt1}
H. Risken, {\em The Fokker-Planck Equation: Methods of Solution and
Applications} (Springer Verlag, Berlin, 1984).

\bibitem{mfpt2}
K. Lindenberg and V.
Seshadri, J. Chem. Phys. {\bf 71}, 4075 (1979); K. Lindenberg
and B. J. West, J. Stat. Phys. {\bf 42}, 201 (1986).

\bibitem{gidas} B. Gidas, W. M. Ni and L. Nirenberg, Comm. Math. Phys.
{\bf 68}, 200-243 (1979).

\bibitem{macarthur} R. H. MacArthur and E. O. Wilson, {\em The Theory of
Island Biogeography} (Princeton Univ. Press, Princeton, NJ, 1967).

\end{thebibliography}
\end{document}